\input amssym.tex
\input epsf
\epsfclipon


\magnification=\magstephalf
\hsize=14.0 true cm
\vsize=19 true cm
\hoffset=1.0 true cm
\voffset=2.0 true cm

\abovedisplayskip=12pt plus 3pt minus 3pt
\belowdisplayskip=12pt plus 3pt minus 3pt
\parindent=1.0em


\font\sixrm=cmr6
\font\eightrm=cmr8
\font\ninerm=cmr9

\font\sixi=cmmi6
\font\eighti=cmmi8
\font\ninei=cmmi9

\font\sixsy=cmsy6
\font\eightsy=cmsy8
\font\ninesy=cmsy9

\font\sixbf=cmbx6
\font\eightbf=cmbx8
\font\ninebf=cmbx9

\font\eightit=cmti8
\font\nineit=cmti9

\font\eightsl=cmsl8
\font\ninesl=cmsl9

\font\sixss=cmss8 at 8 true pt
\font\sevenss=cmss9 at 9 true pt
\font\eightss=cmss8
\font\niness=cmss9
\font\tenss=cmss10

 at 12 true pt
 at 12 true pt
\font\bigrm=cmr10 at 12 true pt
 at 12 true pt
 at 12 true pt

\font\Bigi=cmmi12 at 16 true pt
 at 16 true pt
\font\Bigrm=cmr12 at 16 true pt
 at 16 true pt
 at 16 true pt

\catcode`@=11
\newfam\ssfam

\def\tenpoint{\def\rm{\fam0\tenrm}%
    \textfont0=\tenrm \scriptfont0=\sevenrm \scriptscriptfont0=\fiverm
    \textfont1=\teni  \scriptfont1=\seveni  \scriptscriptfont1=\fivei
    \textfont2=\tensy \scriptfont2=\sevensy \scriptscriptfont2=\fivesy
    \textfont3=\tenex \scriptfont3=\tenex   \scriptscriptfont3=\tenex
    \textfont\itfam=\tenit                  \def\it{\fam\itfam\tenit}%
    \textfont\slfam=\tensl                  \def\sl{\fam\slfam\tensl}%
    \textfont\bffam=\tenbf \scriptfont\bffam=\sevenbf
    \scriptscriptfont\bffam=\fivebf
                                            \def\bf{\fam\bffam\tenbf}%
    \textfont\ssfam=\tenss \scriptfont\ssfam=\sevenss
    \scriptscriptfont\ssfam=\sevenss
                                            \def\ss{\fam\ssfam\tenss}%
    \normalbaselineskip=13pt
    \setbox\strutbox=\hbox{\vrule height8.5pt depth3.5pt width0pt}%
    \let\big=\tenbig
    \normalbaselines\rm}

\def\ninepoint{\def\rm{\fam0\ninerm}%
    \textfont0=\ninerm      \scriptfont0=\sixrm
                            \scriptscriptfont0=\fiverm
    \textfont1=\ninei       \scriptfont1=\sixi
                            \scriptscriptfont1=\fivei
    \textfont2=\ninesy      \scriptfont2=\sixsy
                            \scriptscriptfont2=\fivesy
    \textfont3=\tenex       \scriptfont3=\tenex
                            \scriptscriptfont3=\tenex
    \textfont\itfam=\nineit \def\it{\fam\itfam\nineit}%
    \textfont\slfam=\ninesl \def\sl{\fam\slfam\ninesl}%
    \textfont\bffam=\ninebf \scriptfont\bffam=\sixbf
                            \scriptscriptfont\bffam=\fivebf
                            \def\bf{\fam\bffam\ninebf}%
    \textfont\ssfam=\niness \scriptfont\ssfam=\sixss
                            \scriptscriptfont\ssfam=\sixss
                            \def\ss{\fam\ssfam\niness}%
    \normalbaselineskip=12pt
    \setbox\strutbox=\hbox{\vrule height8.0pt depth3.0pt width0pt}%
    \let\big=\ninebig
    \normalbaselines\rm}

\def\eightpoint{\def\rm{\fam0\eightrm}%
    \textfont0=\eightrm      \scriptfont0=\sixrm
                             \scriptscriptfont0=\fiverm
    \textfont1=\eighti       \scriptfont1=\sixi
                             \scriptscriptfont1=\fivei
    \textfont2=\eightsy      \scriptfont2=\sixsy
                             \scriptscriptfont2=\fivesy
    \textfont3=\tenex        \scriptfont3=\tenex
                             \scriptscriptfont3=\tenex
    \textfont\itfam=\eightit \def\it{\fam\itfam\eightit}%
    \textfont\slfam=\eightsl \def\sl{\fam\slfam\eightsl}%
    \textfont\bffam=\eightbf \scriptfont\bffam=\sixbf
                             \scriptscriptfont\bffam=\fivebf
                             \def\bf{\fam\bffam\eightbf}%
    \textfont\ssfam=\eightss \scriptfont\ssfam=\sixss
                             \scriptscriptfont\ssfam=\sixss
                             \def\ss{\fam\ssfam\eightss}%
    \normalbaselineskip=10pt
    \setbox\strutbox=\hbox{\vrule height7.0pt depth2.0pt width0pt}%
    \let\big=\eightbig
    \normalbaselines\rm}

\def\tenbig#1{{\hbox{$\left#1\vbox to8.5pt{}\right.\n@space$}}}
\def\ninebig#1{{\hbox{$\textfont0=\tenrm\textfont2=\tensy
                       \left#1\vbox to7.25pt{}\right.\n@space$}}}
\def\eightbig#1{{\hbox{$\textfont0=\ninerm\textfont2=\ninesy
                       \left#1\vbox to6.5pt{}\right.\n@space$}}}

\font\sectionfont=cmbx10
\font\subsectionfont=cmti10

\def\figurecaptionfont{\ninepoint}
\def\tablecaptionfont{\ninepoint}
\def\footnotefont{\eightpoint}


\newcount\equationno
\newcount\bibitemno
\newcount\figureno
\newcount\tableno

\equationno=0
\bibitemno=0
\figureno=0
\tableno=0


\footline={\ifnum\pageno=0{\hfil}\else
{\hss\rm\the\pageno\hss}\fi}


\def\section #1. #2 \par
{\vskip0pt plus .10\vsize\penalty-100 \vskip0pt plus-.10\vsize
\vskip 1.6 true cm plus 0.2 true cm minus 0.2 true cm
\global\def\equationlabel{#1}
\global\equationno=0
\leftline{\sectionfont #1. #2}\par
\immediate\write\terminal{Section #1. #2}
\vskip 0.7 true cm plus 0.1 true cm minus 0.1 true cm
\noindent}


\def\subsection #1 \par
{\vskip0pt plus 0.8 true cm\penalty-50 \vskip0pt plus-0.8 true cm
\vskip2.5ex plus 0.1ex minus 0.1ex
\leftline{\subsectionfont #1}\par
\immediate\write\terminal{Subsection #1}
\vskip1.0ex plus 0.1ex minus 0.1ex
\noindent}


\def\appendix #1. #2 \par
{\vskip0pt plus .20\vsize\penalty-100 \vskip0pt plus-.20\vsize
\vskip 1.6 true cm plus 0.2 true cm minus 0.2 true cm
\global\def\equationlabel{\hbox{\rm#1}}
\global\equationno=0
\leftline{\sectionfont Appendix #1. #2}\par
\immediate\write\terminal{Appendix #1. #2}
\vskip 0.7 true cm plus 0.1 true cm minus 0.1 true cm
\noindent}



\def\equation#1{$$\displaylines{\qquad #1}$$}
\def\enum{\global\advance\equationno by 1
\hfill\llap{{\rm(\equationlabel.\the\equationno)}}}

\def\next#1{\cr\noalign{\vskip#1}\qquad}


\def\ifundefined#1{\expandafter\ifx\csname#1\endcsname\relax}

\def\ref#1{\ifundefined{#1}?\immediate\write\terminal{unknown reference
on page \the\pageno}\else\csname#1\endcsname\fi}

\newwrite\terminal
\newwrite\bibitemlist

\def\bibitem#1#2\par{\global\advance\bibitemno by 1
\immediate\write\bibitemlist{\string\def
\expandafter\string\csname#1\endcsname
{\the\bibitemno}}
\item{[\the\bibitemno]}#2\par}

\def\beginbibliography{
\vskip0pt plus .15\vsize\penalty-100 \vskip0pt plus-.15\vsize
\vskip 1.2 true cm plus 0.2 true cm minus 0.2 true cm
\leftline{\sectionfont References}\par
\immediate\write\terminal{References}
\immediate\openout\bibitemlist=biblist
\frenchspacing\parindent=1.8em
\vskip 0.5 true cm plus 0.1 true cm minus 0.1 true cm}

\def\endbibliography{
\immediate\closeout\bibitemlist
\nonfrenchspacing\parindent=1.0em}


\def\figurecaption#1{\global\advance\figureno by 1
\narrower\figurecaptionfont
Fig.~\the\figureno. #1}

\def\tablecaption#1{\global\advance\tableno by 1
\vbox to 0.5 true cm { }
\centerline{\tablecaptionfont%
Table~\the\tableno. #1}
\vskip-0.4 true cm}

\def\thicktablerule{\hrule height1pt}
\def\thintablerule{\hrule height0.4pt}

\tenpoint

\immediate\openin\bibitemlist=biblist
\ifeof\bibitemlist\immediate\closein\bibitemlist
\else\immediate\closein\bibitemlist
\input biblist \fi


\def\thismonth{\ifcase\month\or
January\or February\or March\or April\or May\or June\or
July\or August\or September\or October\or November\or December\fi}



\def\rmd{{\rm d}}
\def\rmD{{\rm D}}
\def\rme{{\rm e}}
\def\rmO{{\rm O}}


\def\Im{{\rm Im}\,}


\def\proof{\noindent{\sl Proof:}\kern0.6em}

\def\frac#1#2{\hbox{$#1\over#2$}}
\def\dual{\mathstrut^*\kern-0.1em}

\def\lvec#1{\setbox0=\hbox{$#1$}
    \setbox1=\hbox{$\scriptstyle\leftarrow$}
    #1\kern-\wd0\smash{
    \raise\ht0\hbox{$\raise1pt\hbox{$\scriptstyle\leftarrow$}$}}
    \kern-\wd1\kern\wd0}
\def\rvec#1{\setbox0=\hbox{$#1$}
    \setbox1=\hbox{$\scriptstyle\rightarrow$}
    #1\kern-\wd0\smash{
    \raise\ht0\hbox{$\raise1pt\hbox{$\scriptstyle\rightarrow$}$}}
    \kern-\wd1\kern\wd0}


\def\nabstar#1{{\nabla\kern0.5pt\smash{\raise 4.5pt\hbox{$\ast$}}
               \kern-5.5pt_{#1}}}

\def\drvstar#1{{\partial\kern0.5pt\smash{\raise 4.5pt\hbox{$\ast$}}
               \kern-6.0pt_{#1}}}

\def\ldrvstar#1{{\lvec{\,\partial}\kern-0.5pt\smash{\raise 4.5pt\hbox{$\ast$}}
               \kern-5.0pt_{#1}}}


\def\MeV{{\rm MeV}}
\def\GeV{{\rm GeV}}

\def\fm{{\rm fm}}




\def\dirac#1{\gamma_{#1}}
\def\diracstar#1#2{
    \setbox0=\hbox{$\gamma$}\setbox1=\hbox{$\gamma_{#1}$}
    \gamma_{#1}\kern-\wd1\kern\wd0
    \smash{\raise4.5pt\hbox{$\scriptstyle#2$}}}


\def\tr{{\rm tr}}

\def\Ad{{\rm Ad}\kern0.1em}



\def\Dhat{\hat{D}}

\def\abar{\bar{a}}

\def\Dmdag{\setbox0=\hbox{$\displaystyle D$}%
           \setbox1=\hbox{$\displaystyle D_m$}%
           D_m\kern-\wd1\kern\wd0%
           \smash{\raise4.5pt\hbox{\kern0pt$\scriptstyle\dagger$}}\kern4pt}



\def\Nsea{N_{\rm sea}}
\def\eps{\epsilon}
\def\MSbar{{\rm\overline{MS\kern-0.05em}\kern0.05em}}

\rightline{CERN-TH/2003-213}
\rightline{CPT-2003/PE.4569}
\rightline{DESY 03-135}
\rightline{MPP-2003-74}

\vskip 1.0 true cm 
\centerline{\Bigrm Lattice QCD in the 
$\textfont0=\Bigrm\textfont1=\Bigi\epsilon$-regime and random
matrix theory}
\vskip 0.6 true cm
\centerline{\bigrm L.~Giusti\footnote{$^{\star}$}{\footnotefont%
On leave from CNRS, Centre de Physique Th\'eorique, F-13288 Marseille, France}%
$^{,1}$,
M.~L\"uscher$^1$, P.~Weisz$^2$, H.~Wittig$^3$}
\vskip1.8ex
\centerline{$^1\hskip-3pt$ \it 
CERN, Theory Division, CH-1211 Geneva 23, Switzerland}
\vskip1.0ex
\centerline{$^2\hskip-3pt$ \it 
Max-Planck-Institut f\"ur Physik, F\"ohringer Ring 6, D-80805 Munich, Germany}
\vskip1.0ex
\centerline{$^3\hskip-3pt$ \it 
DESY, Theory Group, Notkestrasse 85, D-22603 Hamburg, Germany}
\vskip 0.8 true cm
\thintablerule
\vskip 2.0ex
\ninepoint
\leftline{\bf Abstract}
\vskip 1.0ex\noindent
In the $\epsilon$-regime of QCD 
the main features of
the spectrum of the low-lying eigenvalues
of the (euclidean) Dirac operator are expected to be 
described by a certain universality class of random matrix models.
In particular, the latter predict
the joint statistical distribution of 
the individual eigenvalues in any topological sector of the theory.
We compare some of these predictions with 
high-precision numerical data obtained
from lattice QCD for a range of lattice spacings and
volumes.
While no complete matching is observed, 
the results agree with theoretical expectations
at volumes larger than about $5\,\fm^4$.
\vskip 2.0ex
\thintablerule

\tenpoint
\vskip-0.2cm

\section 1. Introduction

The proposition that the low-lying eigenvalues of the Dirac operator
in the so-called $\eps$-regime of QCD
[\ref{GasserLeutwyler},\ref{LeutwylerSmilga}]
are distributed in the same way as the eigenvalues of a large random matrix
was put forward a number of years ago
[\ref{ShuryakVerbaarschot}--\ref{Verbaarschot}]
and has since then been worked out in great detail
(see refs.~[\ref{VerbaarschotReview},\ref{DamgaardReview}] 
for a review and further references).
Perhaps the most important qualitative prediction of random matrix 
theory is that, at vanishing quark masses, the
eigenvalues scale proportionally to $(\Sigma V)^{-1}$,
where $\Sigma$ denotes the (bare) quark condensate 
and $V$ the space-time volume.
In particular,
the spectrum near the origin rapidly becomes very dense 
when the volume increases.

So far the correctness of the proposition
has not been established from first principles.
Apart from symmetry and universality arguments,
it is supported by 
chiral perturbation theory,
where the Leutwyler--Smilga sum rules 
[\ref{LeutwylerSmilga}] can be shown
to be reproduced, at large volumes, by random matrix theory.
To date there is, however, no similar theoretical check on the 
distributions of the individual eigenvalues
[\ref{NishigakiDamgaardWettig},\ref{DamgaardNishigaki}].

Random matrix theory is well defined for any number $\Nsea\geq0$
of sea quarks. It is thus possible to extend the proposition
to quenched QCD, even though the latter tends to be
singular in the chiral limit.
Another instance where random matrix theory
with $\Nsea=0$ may be expected to apply 
is full QCD with quark masses $m$ such that $m\Sigma V\gg1$.
The quark determinant is safely bounded from 
below in this case, and the distributions of the 
low-lying eigenvalues of the {\it massless} Dirac operator 
should consequently
be as in quenched QCD (up to a scale transformation).

The present paper is part of an ongoing project
whose final goal is to extract physical
parameters, such as the pion decay constant 
and the electroweak effective couplings,
from numerical simulations of
lattice QCD in the $\eps$-regime
[\ref{NumMethods},\ref{NextPaper}].
Since chiral symmetry plays a central r\^ole in this context,
we use a lattice Dirac operator that satisfies 
the Ginsparg--Wilson relation and thus preserves the symmetry
[\ref{GinspargWilson}--\ref{Locality}].
Unfortunately numerical simulations of
full QCD then become even more demanding
than they normally are,
and in this paper we shall, therefore, only consider the case
of quenched QCD.

Comparisons of random matrix theory with simulation data
obtained from lattice QCD with exact chiral symmetry have 
previously been published by a number of collaborations 
[\ref{EdwardsHellerKiskisNarayanan}--\ref{BietenholzEtAl}].
Here we extend these studies to significantly larger lattices
and collect data at different lattice spacings so as 
to be able to check for lattice effects.
We wish to add that simulations in the $\eps$-regime of QCD 
become increasingly difficult at large volumes, 
because the low-lying eigenvalues of the Dirac operator are
extremely small and closely spaced. The use 
of efficient techniques such as those 
described in ref.~[\ref{NumMethods}] is thus essential.

\section 2. Random matrix model

To set up notations, we now briefly recall the definition of the matrix
model that is expected to describe the spectrum of the 
Dirac operator in QCD. There is actually a whole universality
class of such models, and we will only describe the simplest
representative, the so-called gaussian chiral unitary model.

Let us consider $N\times N$ matrices of the form 
\equation{
  \Dhat=\pmatrix{0 & W\cr -W^{\dagger} &0\cr}\!\!
  {\left.\right\}N_{+}\atop\left.\right\}N_{-}},
  \qquad N=N_{+}+N_{-}, 
  \enum
}
where $W$ is a complex rectangular random matrix.
We use the symbol $\Dhat$ to indicate that the matrix represents
the massless Dirac operator in the matrix model. 
It is then natural to think of $N$ as the space-time volume
(times some proportionality constant), while
the block structure of $\Dhat$ is
interpreted as a chiral decomposition.
Moreover, since any matrix of this form has $|N_{+}-N_{-}|$ chiral 
zero modes, the index
\equation{
  \nu=N_{+}-N_{-}
  \enum
}
may be identified with the topological charge in QCD.

The observables $\cal O$ in the matrix model are arbitrary 
functions of matrix elements of $W$, and we 
are interested in the large-$N$ limit of the 
expectation values
\equation{
  \quad\langle{\cal O}\rangle_{\nu}={1\over{\cal Z}_{\nu}}
  \int\rmD[W]\,{\cal O}\det(\Dhat+m)^{\Nsea}\,
  \rme^{-\frac{1}{2}N\tr\left\{W^{\dagger}W\right\}}
  \enum
}
at fixed $\mu=mN$. 
Clearly $m$ represents the quark mass in the matrix model
(in the case where all quarks have the same mass), but 
this parameter is actually irrelevant in the present paper, since
we shall compare the matrix model with quenched QCD and thus set 
$\Nsea=0$ from the beginning.

Apart from the chiral zero modes, all eigenvalues of $\Dhat$
come in complex conjugate pairs $\pm i\lambda_k$ 
that can be ordered according to
\equation{
   0\leq\lambda_1\leq\lambda_2\leq\ldots
   \enum
} 
The expectation values of these eigenvalues
scale proportionally to $1/N$.
In particular, the spectral densities at fixed topology,
\equation{
   \rho_{k,\nu}(z)=
   \lim_{N\to\infty}
   \left\langle\delta(z-\lambda_kN)\right\rangle_{\nu},
   \qquad
   0\leq z<\infty,
   \enum
}
as well as the corresponding joint distributions
of the first $n$ eigenvalues
are well defined and analytically calculable 
[\ref{NishigakiDamgaardWettig},\ref{DamgaardNishigaki}].
Using these expressions, 
the expectation values of the first few eigenvalues and of 
their products can be obtained exactly, although in practice 
the formulae rapidly become
so complicated that numerical methods
are required.

\section 3. Numerical simulation

We assume that the reader is familiar with the standard formulations of
lattice QCD and the Ginsparg--Wilson relation
as discussed in refs.~[\ref{Hasenfratz}--\ref{Locality}], for example.
In this section we summarize the parameter choices
that we have made and discuss what precisely is being 
compared with random matrix theory.

\subsection 3.1 Lattice theory

We consider four-dimensional lattices of size $L$ in all
dimensions and impose periodic boundary conditions on the fields.
The action of the SU(3) gauge field is taken to be the standard
plaquette action with bare coupling $g_0$.
As already mentioned, 
the lattice Dirac operator $D$ should satisfy the Ginsparg--Wilson
relation, 
\equation{
   \dirac{5}D+D\dirac{5}=\abar D\dirac{5}D,
   \enum
}
in addition to the usual requirements such as locality.
For this study we decided to use the
Neuberger--Dirac operator [\ref{NeubergerDirac}],
with shift parameter $s=0.4$ [\ref{Locality}], in view of its 
relative simplicity.
The parameter $\abar$ in eq.~(3.1) is then 
given by
\equation{
  \abar=a/(1+s),
  \enum
}
where $a$ denotes the lattice spacing
(our notational conventions are as in
refs.~[\ref{NumMethods},\ref{Locality}]). 

As usual we define the index $\nu$
of $D$ to be the difference $n_{+}-n_{-}$ of 
the numbers of exact zero modes with 
positive and negative chirality.
A well-known implication of the Ginsparg--Wilson relation 
is that $\nu=Q$, where 
\equation{
  Q=a^4\sum_xq(x),
  \qquad
  q(x)=-\frac{1}{2}\abar\,\tr\{\dirac{5}D(x,x)\},
  \enum
}
can be taken as
the definition of the topological charge of the gauge field
[\ref{HLN}]. In particular, the charge density $q(x)$ is a local
gauge-invariant expression in the link variables 
[\ref{Locality}].

\subsection 3.2 Numerical techniques

At large volumes the low-lying non-zero eigenvalues 
of $D$ are orders of magnitude smaller than the largest eigenvalues.
It is hence important to choose a numerical im\-ple\-mentation 
of the Neuberger--Dirac operator, where the approximation
errors can be guaranteed to be sufficiently small for the spectrum
to be obtained to the required level of precision.

The minmax polynomial approximation
that was introduced in ref.~[\ref{NumMethods}]
provides a solution to this problem.
We also apply some of the other numerical techniques mentioned there.
In particular, the eigenvalues of $D$ are obtained by minimization
of the Ritz functional [\ref{EvaI},\ref{EvaII}] of the hermitian operators
\equation{
   D^{\pm}=P_{\pm}DP_{\pm},
   \qquad
   P_{\pm}=\frac{1}{2}(1\pm\dirac{5}),
   \enum
}
in the positive and negative chirality sectors.

\subsection 3.3 Simulation parameters

We have simulated altogether $7$ lattices with various sizes
and lattice spacings (see table~1).
There are three groups of lattices, labelled A, B and C,
where the lattice size $L$ is kept fixed in physical units.
Within each of these groups, the 
bare coupling $\beta=6/g_0^2$ 
is thus the only parameter that varies, which allows us
to obtain a direct check on the dependence of the 
calculated observables on the lattice spacing.
For the conversion to physical units we use 
the recent parametrization of the Sommer scale
$r_0$ [\ref{SommerScaleA}]
by the ALPHA collaboration (eq.~(2.6) of [\ref{SommerScaleB}])
and set $r_0=0.5\,\fm$.

\topinsert
\newdimen\digitwidth
\setbox0=\hbox{\rm 0}
\digitwidth=\wd0
\catcode`@=\active
\def@{\kern\digitwidth}
\tablecaption{Simulation parameters}
\vskip1.0ex
$$\vbox{\settabs\+&%
                  xxxxxxx&&
                  xxxxxxx&&
                  xxxxx&&
                  xxxxxxx&&
                  xxxxxxxx&&
                  xxxxxx&&
                  x&\cr
\thicktablerule
\vskip1ex
                \+& \hfill lattice \hfill
                 && \hfill $\beta$\hfill
                 && \hfill $L/a$\hfill
                 && \hfill $r_0/a$\hfill
                 && \hfill $L$ [fm]\hfill
                 && \hfill $N_{\rm meas}$\hfill
                 && \cr
\vskip1.0ex
\thintablerule
\vskip1.5ex
  \+& \hfill ${\rm A}_1$\hfill
  &&  \hskip1ex $6.0$\hfill 
  &&  \hfill $12$\hfill
  &&  \hfill $5.368$\hfill
  &&  \hfill $1.12$\hfill 
  &&  \hfill $2452$\hskip1ex
  &\cr
\vskip0.3ex
  \+& \hfill ${\rm A}_2$\hfill
  &&  \hskip1ex $6.1791$\hfill 
  &&  \hfill $16$\hfill
  &&  \hfill $7.158$\hfill
  &&  \hfill $1.12$\hfill 
  &&  \hfill $1138$\hskip1ex
  &\cr
\vskip0.3ex
  \+& \hfill ${\rm B}_0$\hfill
  &&  \hskip1ex $5.8458$\hfill 
  &&  \hfill $12$\hfill
  &&  \hfill $4.026$\hfill
  &&  \hfill $1.49$\hfill 
  &&  \hfill $2918$\hskip1ex
  &\cr
\vskip0.3ex
  \+& \hfill ${\rm B}_1$\hfill
  &&  \hskip1ex $6.0$\hfill 
  &&  \hfill $16$\hfill
  &&  \hfill $5.368$\hfill
  &&  \hfill $1.49$\hfill 
  &&  \hfill $1001$\hskip1ex
  &\cr
\vskip0.3ex
  \+& \hfill ${\rm B}_2$\hfill
  &&  \hskip1ex $6.1366$\hfill 
  &&  \hfill $20$\hfill
  &&  \hfill $6.710$\hfill
  &&  \hfill $1.49$\hfill 
  &&  \hfill $963$\hskip1ex
  &\cr
\vskip0.3ex
  \+& \hfill ${\rm C}_0$\hfill
  &&  \hskip1ex $5.8784$\hfill 
  &&  \hfill $16$\hfill
  &&  \hfill $4.294$\hfill
  &&  \hfill $1.86$\hfill 
  &&  \hfill $1109$\hskip1ex
  &\cr
\vskip0.3ex
  \+& \hfill ${\rm C}_1$\hfill
  &&  \hskip1ex $6.0$\hfill 
  &&  \hfill $20$\hfill
  &&  \hfill $5.368$\hfill
  &&  \hfill $1.86$\hfill 
  &&  \hfill $931$\hskip1ex
  &\cr
\vskip1.0ex
\thicktablerule
}$$
\endinsert

In all these simulations the generation of the gauge-field 
configurations consumes a negligible amount of computer time.
We have therefore performed many update cycles
between subsequent configurations (typically $500$
iterations of $1$ heatbath and at least $6$
over-relaxation updates of all link variables)
so that they can be assumed to be statistically independent.
The number of gauge fields in each ensemble 
quoted in the last column of table~1
coincides with the number of ``measurements" of the 
topological charge and the 
lowest few eigenvalues of the Dirac operator.

\subsection 3.4 Matching with random matrix theory

The eigenvalues $\gamma$ of $D$ lie on a circle
in the complex plane,
\equation{
  D\psi=\gamma\psi,
  \qquad
  \gamma={1\over\abar}\left(1-\rme^{i\phi}\right),
  \enum
}
and come in complex conjugate pairs (if\/ $\Im\gamma$ does not vanish).
We cannot directly compare these eigenvalues
with the eigenvalues of the random matrix $\Dhat$ 
since the latter lie on the imaginary axis.
The important point to note is, however, that
the radius of the circle diverges 
in the continuum limit and that the real parts of the
eigenvalues $\gamma$ with $|\gamma|\ll1/\abar$ rapidly go to zero
in this limit.
In practice we set (for these eigenvalues)
\equation{
  \lambda=|\gamma|=
  {1\over\abar}\left\{2\left(1-\cos\phi\right)\right\}^{1/2}
  \enum
}
and
compare the distributions of the scaled eigenvalues
\equation{
  z=\lambda\Sigma V,
  \qquad
  V=L^4,
  \enum
}
with those of the scaled eigenvalues $z=\lambda N$ in the matrix model.
Evidently the definition (3.6) is arbitrary to some extent,
but other suggested formulae differ 
by terms of order $\abar^2\lambda^3$ and are thus asymptotically
equivalent. In particular, for the analysis of our data this ambiguity
turns out to be numerically insignificant.

Although we use the same symbol for 
the proportionality constant
$\Sigma$ in eq.~(3.7) as for the quark condensate in full QCD,
it should be emphasized that in the present context 
$\Sigma$ is just a free parameter with no obvious physical 
interpretation.
The quark condensate is actually not even 
well defined in quenched QCD, since it is not possible to 
pass to the chiral limit in this theory.
We thus prefer to regard $\Sigma$ as an effective parameter that is to be
determined, on any given lattice, by comparing the spectral 
distributions of $D$ with those of $\Dhat$.

\section 4. Distribution of the topological charge

We now first discuss the probability 
$P_{\nu}$ to find a gauge field with topological charge $Q=\nu$.
Random matrix theory does not provide any 
information on $P_{\nu}$, and we are thus uniquely
concerned with properties of QCD in this section.
Basically we wish to find out whether the charge distribution
scales in the expected way as a function of the volume and 
the lattice spacing. If this is the case, it will then 
be more plausible that the eigenvalues
of the Dirac operator behave coherently in the different sectors
(as suggested by random matrix theory).

\subsection 4.1 Large volume limit

The fact that the topological charge $Q$
is given in terms of a local density $q(x)$
allows us to derive an asymptotic formula for 
$P_{\nu}$ at large volumes. First note that since $Q$ is integer-valued,
we have%
\kern1.5pt\footnote{$\dag$}{\footnotefont%
As in the matrix model, expectation values in QCD 
at fixed charge $Q=\nu$ are denoted by 
$\langle\ldots\rangle_{\nu}$. Brackets without lower index
imply an unconstrained expectation value and 
$\langle\phi_1(x_1)\ldots\phi_n(x_n)\rangle^{\rm con}$ 
stands for the connected part of the full $n$-point correlation function 
of the fields $\phi_1,\ldots,\phi_n$.}
\equation{
  P_{\nu}=\int_{-\pi}^{\pi}{\rmd\theta\over2\pi}\,\rme^{-i\theta\nu}
  \rme^{-F(\theta)},
  \qquad
  F(\theta)=-\ln\left\{\left\langle\rme^{i\theta Q}\right\rangle\right\}.
  \enum
}
Using the moment-cumulant transformation, the free energy $F(\theta)$
may then be expanded in a series
\equation{
  F(\theta)=V\sum_{n=1}^{\infty}(-1)^{n+1}{\theta^{2n}\over(2n)!}C_n,
  \enum
  \next{2ex}
  C_n=a^{8n-4}\sum_{x_1,\ldots,x_{2n-1}}
  \left\langle q(x_1)\ldots q(x_{2n-1})q(0)\right\rangle^{\rm con}.
  \enum
}
In quenched QCD the connected correlation functions in this formula
are evaluated in the pure gauge theory, where all particles are
fairly heavy (the mass of the lightest glueball is around $1.6\,\GeV$).
As a result the cumulants $C_n$ approach their infinite-volume
limit exponentially fast, i.e.~as soon as $L$ is larger than
$1\,\fm$ or so, they can be expected to be practically independent 
of the volume.

If we now insert eq.~(4.2) in eq.~(4.1), it
is clear that the integral is dominated by the saddle point at $\theta=0$
in the large-volume limit. 
The asymptotic formula
\equation{
  P_{\nu}={\rme^{-{\nu^2\over2\sigma^2}}\over\sqrt{2\pi\sigma^2}}
  \left\{1+\rmO\left(V^{-1}\right)\right\}
  \enum
}
is thus obtained, where $\sigma^2=\langle Q^2\rangle$.
On the lattices ${\rm A}_1,\ldots,{\rm C}_1$, the 
observed charge distributions are in fact statistically 
consistent with the leading term in eq.~(4.4) (see fig.~1).
From this point of view the lattices are hence 
in the large-volume regime.

\topinsert
\vbox{
\vskip0.0cm
\epsfxsize=8.4cm\hskip1.8cm\epsfbox{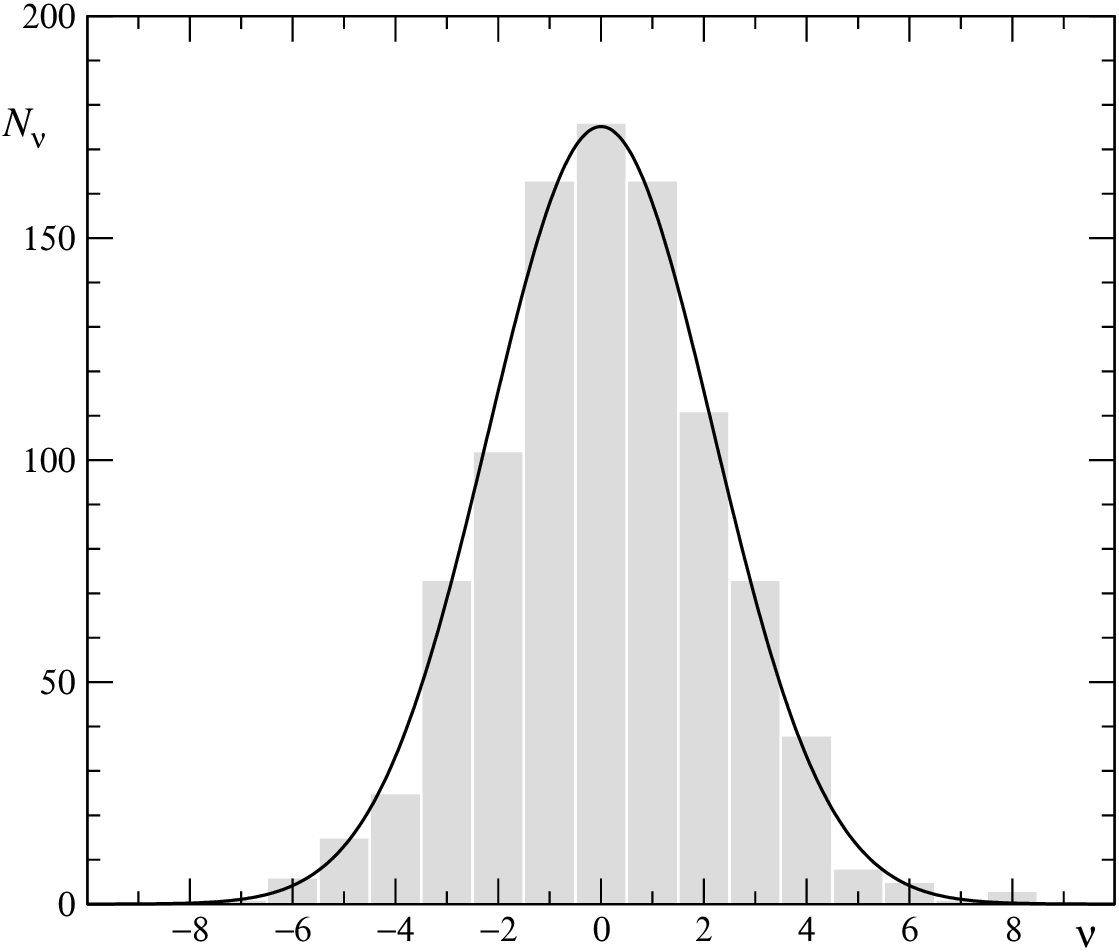}
\vskip0.4cm
\figurecaption{%
Histogram of the topological charge on lattice ${\rm B}_2$.
The curve represents the leading term in the large-volume 
formula~(4.4)
with $\sigma^2=\langle Q^2\rangle$ taken from table~2.
}
\vskip0.0cm
}
\endinsert

\subsection 4.2 Determination of the topological susceptibility

The width of the charge distribution (or, equivalently,
the susceptibility $\chi=\langle Q^2\rangle/V$) may
be calculated straightforwardly by averaging 
$Q^2$ over the ensemble of gauge configurations that was
generated. However, this procedure may not be safe in general,
because the tails of the charge distribution are poorly sampled.

To understand what the problem is, first note that 
the number $N_{\nu}$ of configurations with
charge $Q=\nu$ in a sample of $N$ confi\-gu\-rations is distributed
according to Poisson statistics.
In particular, the variance of $N_{\nu}$ is equal to its mean
value $NP_{\nu}$.
If $N_{\nu}$ is small,
the empirical probability $N_{\nu}/N$ may consequently 
be quite different from the true probability $P_{\nu}$.
We thus compute $\langle Q^2\rangle$ as
a sum of two contributions,
\equation{
  \sum_{|\nu|\leq\nu_{\rm max}}
  \nu^2\, {N_{\nu}\over N}+\sum_{|\nu|>\nu_{\rm max}}\nu^2P_{\nu},
  \enum
}
where $\nu_{\max}$ is the maximal charge such that
$N_{\nu}+N_{-\nu}$ is at least $10$ for all $\nu$ in the first sum.
To evaluate the second sum, we insert the large-volume expression
(4.4) for the exact distribution $P_{\nu}$, setting $\sigma^2$ to the
naive estimate of $\langle Q^2\rangle$, where
any exceptional configurations with absolute charges 
$|Q|$ significantly
larger than $\nu_{\max}$
should be discarded (there were none in our samples).
The second sum is actually a small correction to the first, and
other estimates of $\sigma^2$ could therefore be used at this point 
with little effect on the results listed in table~2.\kern1pt%
\footnote{$\dagger$}{\footnotefont%
To the extent they can be compared, the results obtained
in a similar study of the topological susceptibility 
by Del Debbio and Pica [\ref{LuigiClaudio}], 
which appeared during the completion of the present paper,
are compatible with those reported here.}

\topinsert
\newdimen\digitwidth
\setbox0=\hbox{\rm 0}
\digitwidth=\wd0
\catcode`@=\active
\def@{\kern\digitwidth}
\tablecaption{Topological susceptibility}
\vskip-0.5ex
$$\vbox{\settabs\+&%
                  xxxxx&&%
                  xxxxxxxxx&&
                  xxxxxxxxx&&
                  xxxxxxxxx&&
                  xxxxxxxxx&&
                  xxxxxxxxx&&
                  xxxxxxxxx&&
                  xxxxxxxxx&&
                  &\cr
\thicktablerule
\vskip1ex
                \+& 
                 && \hfill ${\rm A}_1$\hfill
                 && \hfill ${\rm A}_2$\hfill
                 && \hfill ${\rm B}_0$\hfill
                 && \hfill ${\rm B}_1$\hfill
                 && \hfill ${\rm B}_2$\hfill
                 && \hfill ${\rm C}_0$\hfill
                 && \hfill ${\rm C}_1$\hfill
                 & \cr
\vskip1.0ex
\thintablerule
\vskip1.5ex
  \+& \hfill $\langle Q^2\rangle $\hfill
  &&  \hfill  $1.63(5)$\hfill
  &&  \hfill  $1.59(8)$\hfill
  &&  \hfill  $5.6(2)$\hfill
  &&  \hfill  $5.6(3)$\hfill
  &&  \hfill  $4.8(2)$\hfill
  &&  \hfill  $15.0(7)$\hfill
  &&  \hfill  $12.8(9)$\hfill
  &\cr
\vskip0.5ex
  \+& \hfill  $\chi r_0^4$\hfill
  &&  \hfill  $@0.065(2)$ \hfill
  &&  \hfill  $@0.064(3)$ \hfill
  &&  \hfill  $@0.071(2)$ \hfill
  &&  \hfill  $@0.071(4)$ \hfill
  &&  \hfill  $@0.061(3)$ \hfill
  &&  \hfill  $@0.078(4)$ \hfill
  &&  \hfill  $@0.066(5)$ \hfill
  &\cr
\vskip1.0ex
\thicktablerule
\vskip1.5ex
\+&\kern1pt%
{\footnotefont\noindent%
The errors quoted in the second row 
do not include the error on $r_0/a$}&\cr
}$$
\vskip-2.0ex
\endinsert

The fact that $\chi r_0^4$ comes out to be the same within errors 
on the lattices ${\rm A}_1,{\rm B}_1$ and ${\rm C}_1$
(which have exactly the same lattice spacing but different volumes)
is in line with the above argumentation
that finite-volume effects in
connected correlation functions
of local fields should be small in the pure gauge theory
when $L\geq1\,\fm$.
As far as the volume dependence is
concerned, the charge distribution on the lattices 
that we have simulated thus behaves entirely according 
to expectations.

\subsection 4.3 Continuum limit

It is often taken for granted that 
the division of 
the space of gauge fields into topological sectors
and the topological susceptibility
have a well-defined meaning in the continuum limit of QCD.
Beyond the semi-classical approximation, where the functional
integral is expanded about the instanton solutions, the
question remains undecided, however, and is in fact
difficult to pose in precise terms 
without reference to a regularization of the theory.
In particular, the susceptibility cannot simply be defined
as the integral over the two-point function
$\langle q(x)q(0)\rangle$ of the charge density, because
there is a non-integrable singularity at $x=0$. 

In lattice QCD the space of fields is connected and the
assignment of a topological charge to every
lattice gauge field is consequently not unique.
The definition (3.3), for example,
depends on the choice of $D$.
Close to the continuum limit, the integration measure 
in the functional integral may, however, be
increasingly supported on fields where the charge assignment
is unambiguous. The topological sectors would then 
arise dynamically and any differences in the definition
of the charge would be ultimately irrelevant.

This picture suggests that 
the susceptibility should behave like a physical quantity 
of dimension $4$, i.e.~that the combination $\chi r_0^4$ should 
be independent of the lattice spacing, up to corrections of 
order $a^2$. The data listed in the second row of table~2 may actually 
be fitted by 
the linear expression $\chi r_0^4=c_0+c_1a^2$,
and our results are, therefore, statistically consistent with
the existence of a well-defined continuum limit of the 
topological susceptibility. If we fit the data
from the B lattices only, taking the error on $r_0/a$ 
into account [\ref{SommerScaleB}],
the value $\chi r_0^4=0.059(5)$ is
obtained at $a=0$, but this number should evidently be used
with caution, as it is determined by linear extrapolation
in $a^2$
of only three data points.

\section 5. Comparison with random matrix theory

As discussed in subsect.~3.4, the matching of the QCD spectra with
random matrix theory involves an unknown scale $\Sigma$
that must be determined from the data.
This complication can be avoided by considering 
ratios of expectation values,
and we now look at some of these first.

\subsection 5.1 Scale-independent tests

In table~3 we list our results for the ratios 
$\langle\lambda_k\rangle_{\nu}/\langle\lambda_j\rangle_{\nu}$
on the lattices ${\rm A}_1,\ldots,{\rm C}_1$.
There is a significant cancellation of statistical errors 
when the ratios are formed, particularly 
for the larger values of $k$ and $j$.
For this reason we show the data for all
$1\leq j<k\leq4$, even though 
some of them are related to each other.
The figures in the last column of table~3 are the values of the
ratios predicted by random matrix theory 
[\ref{NishigakiDamgaardWettig},\ref{DamgaardNishigaki}].

\topinsert
\newdimen\digitwidth
\setbox0=\hbox{\rm 0}
\digitwidth=\wd0
\catcode`@=\active
\def@{\kern\digitwidth}
\tablecaption{Simulation results for the ratios 
$\langle\lambda_k\rangle_{\nu}\kern1pt/\kern1pt
\langle\lambda_j\rangle_{\nu}$}
\vskip0.0ex
$$\vbox{\settabs\+&%
                  xxxi&&
                  xxxxa&&
                  xxxxxxxi&&
                  xxxxxxxi&&
                  xxxxxxxa&&
                  xxxxxxxa&&
                  xxxxxxxa&&
                  xxxxxxxa&&
                  xxxxxxxa&&
                  xxxxii&&
                  x&\cr
\thicktablerule
\vskip1ex
                \+& \hfill $\nu$\hfill
                 && \hfill $k/j$\hfill
                 && \hfill ${\rm A}_1$\hfill
                 && \hfill ${\rm A}_2$\hfill
                 && \hfill ${\rm B}_0$\hfill
                 && \hfill ${\rm B}_1$\hfill
                 && \hfill ${\rm B}_2$\hfill
                 && \hfill ${\rm C}_0$\hfill
                 && \hfill ${\rm C}_1$\hfill
                 && \hfill ${\rm RMT}$\hfill
                 && \cr
\vskip1.0ex
\thintablerule
\vskip1.5ex
  \+& \hfill$0$\hfill
  &&  \hfill$2/1$\hfill 
  &&  \hfill$2.29(4)$\hfill
  &&  $@2.28(6)$\hskip1ex
  &&  $@2.71(6)$\hskip1ex
  &&  $@2.73(10)$\hfill
  &&  $@2.56(10)$\hfill
  &&  $@2.77(12)$\hfill 
  &&  $@3.01(14)$\hfill
  &&  $@2.70$\hfill
  &\cr
\vskip0.3ex
  \+& \hfill$0$\hfill
  &&  \hfill$3/1$\hfill 
  &&  \hfill$3.29(7)$\hfill
  &&  $@3.25(9)$\hskip1ex
  &&  $@4.45(11)$\hfill 
  &&  $@4.59(18)$\hfill
  &&  $@4.43(18)$\hfill
  &&  $@4.65(22)$\hfill 
  &&  $@4.89(23)$\hfill
  &&  $@4.46$\hfill
  &\cr
\vskip0.3ex
  \+& \hfill$0$\hfill
  &&  \hfill$4/1$\hfill 
  &&  \hfill$4.07(8)$\hfill
  &&  $@4.00(11)$\hfill
  &&  $@6.25(15)$\hfill 
  &&  $@6.55(26)$\hfill
  &&  $@6.15(26)$\hfill
  &&  $@6.67(31)$\hfill 
  &&  $@6.99(34)$\hfill
  &&  $@6.22$\hfill
  &\cr
\vskip0.3ex
  \+& \hfill$0$\hfill
  &&  \hfill$3/2$\hfill 
  &&  \hfill$1.44(1)$\hfill
  &&  $@1.42(2)$\hskip1ex
  &&  $@1.65(2)$\hskip1ex
  &&  $@1.68(3)$\hskip1ex
  &&  $@1.73(3)$\hskip1ex
  &&  $@1.68(4)$\hskip1ex
  &&  $@1.63(4)$\hfill
  &&  $@1.65$\hfill
  &\cr
\vskip0.3ex
  \+& \hfill$0$\hfill
  &&  \hfill$4/2$\hfill 
  &&  \hfill$1.78(2)$\hfill
  &&  $@1.75(2)$\hskip1ex
  &&  $@2.31(3)$\hskip1ex
  &&  $@2.40(5)$\hskip1ex
  &&  $@2.40(5)$\hskip1ex
  &&  $@2.41(6)$\hskip1ex
  &&  $@2.32(6)$\hfill
  &&  $@2.30$\hfill
  &\cr
\vskip0.3ex
  \+& \hfill$0$\hfill
  &&  \hfill$4/3$\hfill 
  &&  \hfill$1.24(1)$\hfill
  &&  $@1.23(1)$\hskip1ex
  &&  $@1.40(1)$\hskip1ex
  &&  $@1.43(2)$\hskip1ex
  &&  $@1.39(2)$\hskip1ex
  &&  $@1.44(2)$\hskip1ex
  &&  $@1.43(2)$\hfill
  &&  $@1.40$\hfill
  &\cr
\vskip1.0ex
\thintablerule
\vskip1.0ex
  \+& \hfill$1$\hfill
  &&  \hfill$2/1$\hfill 
  &&  \hfill$1.78(2)$\hfill
  &&  $@1.73(3)$\hskip1ex
  &&  $@2.04(3)$\hskip1ex
  &&  $@2.04(4)$\hskip1ex
  &&  $@2.12(4)$\hskip1ex
  &&  $@2.03(5)$\hskip1ex
  &&  $@1.95(5)$\hfill
  &&  $@2.02$\hfill
  &\cr
\vskip0.3ex
  \+& \hfill$1$\hfill
  &&  \hfill$3/1$\hfill 
  &&  \hfill$2.35(3)$\hfill
  &&  $@2.27(4)$\hskip1ex
  &&  $@3.08(4)$\hskip1ex
  &&  $@3.08(6)$\hskip1ex
  &&  $@3.23(7)$\hskip1ex
  &&  $@3.13(8)$\hskip1ex
  &&  $@2.97(7)$\hfill
  &&  $@3.03$\hfill
  &\cr
\vskip0.3ex
  \+& \hfill$1$\hfill
  &&  \hfill$4/1$\hfill 
  &&  \hfill$2.80(3)$\hfill
  &&  $@2.69(4)$\hskip1ex
  &&  $@4.06(6)$\hskip1ex
  &&  $@4.08(9)$\hskip1ex
  &&  $@4.30(9)$\hskip1ex
  &&  $@4.27(11)$\hfill 
  &&  $@4.15(10)$\hfill
  &&  $@4.04$\hfill
  &\cr
\vskip0.3ex
  \+& \hfill$1$\hfill
  &&  \hfill$3/2$\hfill 
  &&  \hfill$1.32(1)$\hfill
  &&  $@1.31(1)$\hskip1ex
  &&  $@1.51(1)$\hskip1ex
  &&  $@1.51(2)$\hskip1ex
  &&  $@1.53(2)$\hskip1ex
  &   $@1.55(2)$\hskip1ex
  &&  $@1.53(2)$\hfill
  &&  $@1.50$\hfill
  &\cr
\vskip0.3ex
  \+& \hfill$1$\hfill
  &&  \hfill$4/2$\hfill 
  &&  \hfill$1.57(1)$\hfill
  &&  $@1.56(1)$\hskip1ex
  &&  $@1.99(2)$\hskip1ex
  &&  $@2.00(3)$\hskip1ex
  &&  $@2.03(3)$\hskip1ex
  &&  $@2.11(3)$\hskip1ex
  &&  $@2.13(4)$\hfill
  &&  $@2.00$\hfill
  &\cr
\vskip0.3ex
  \+& \hfill$1$\hfill
  &&  \hfill$4/3$\hfill 
  &&  \hfill$1.19(1)$\hfill
  &&  $@1.19(1)$\hskip1ex
  &&  $@1.32(1)$\hskip1ex
  &&  $@1.32(1)$\hskip1ex
  &&  $@1.33(1)$\hskip1ex
  &&  $@1.36(2)$\hskip1ex
  &&  $@1.40(2)$\hfill
  &&  $@1.33$\hfill
  &\cr
\vskip1.0ex
\thintablerule
\vskip1.0ex
  \+& \hfill$2$\hfill
  &&  \hfill$2/1$\hfill 
  &&  \hfill$1.55(2)$\hfill
  &&  $@1.55(3)$\hskip1ex
  &&  $@1.80(2)$\hskip1ex
  &&  $@1.83(4)$\hskip1ex
  &&  $@1.80(4)$\hskip1ex
  &&  $@1.84(4)$\hskip1ex
  &&  $@1.82(3)$\hfill
  &&  $@1.76$\hfill
  &\cr
\vskip0.3ex
  \+& \hfill$2$\hfill
  &&  \hfill$3/1$\hfill 
  &&  \hfill$1.98(3)$\hfill
  &&  $@1.95(4)$\hskip1ex
  &&  $@2.55(3)$\hskip1ex
  &&  $@2.60(6)$\hskip1ex
  &&  $@2.51(6)$\hskip1ex
  &&  $@2.66(6)$\hskip1ex
  &&  $@2.66(5)$\hfill
  &&  $@2.50$\hfill
  &\cr
\vskip0.3ex
  \+& \hfill$2$\hfill
  &&  \hfill$4/1$\hfill 
  &&  \hfill$2.28(3)$\hfill
  &&  $@2.27(5)$\hskip1ex
  &&  $@3.26(4)$\hskip1ex
  &&  $@3.30(8)$\hskip1ex
  &&  $@3.23(8)$\hskip1ex
  &&  $@3.50(8)$\hskip1ex
  &&  $@3.51(7)$\hfill
  &&  $@3.24$\hfill
  &\cr
\vskip0.3ex
  \+& \hfill$2$\hfill
  &&  \hfill$3/2$\hfill 
  &&  \hfill$1.28(1)$\hfill
  &&  $@1.26(1)$\hskip1ex
  &&  $@1.42(1)$\hskip1ex
  &&  $@1.42(2)$\hskip1ex
  &&  $@1.39(2)$\hskip1ex
  &&  $@1.44(2)$\hskip1ex
  &&  $@1.46(2)$\hfill
  &&  $@1.42$\hfill
  &\cr
\vskip0.3ex
  \+& \hfill$2$\hfill
  &&  \hfill$4/2$\hfill 
  &&  \hfill$1.48(1)$\hfill
  &&  $@1.47(2)$\hskip1ex
  &&  $@1.82(2)$\hskip1ex
  &&  $@1.80(3)$\hskip1ex
  &&  $@1.79(3)$\hskip1ex
  &&  $@1.90(3)$\hskip1ex
  &&  $@1.93(3)$\hfill
  &&  $@1.83$\hfill
  &\cr
\vskip0.3ex
  \+& \hfill$2$\hfill
  &&  \hfill$4/3$\hfill 
  &&  \hfill$1.16(1)$\hfill
  &&  $@1.16(1)$\hskip1ex
  &&  $@1.28(1)$\hskip1ex
  &&  $@1.27(1)$\hskip1ex
  &&  $@1.29(1)$\hskip1ex
  &&  $@1.32(1)$\hskip1ex
  &&  $@1.32(1)$\hfill
  &&  $@1.29$\hfill
  &\cr
\vskip1.0ex
\thicktablerule
}$$
\vskip-2ex plus 2ex
\endinsert

A number
of observations can be made at this point:

\vskip1ex minus 0.2ex
\noindent
(1)~For fixed physical volume (i.e.~within the groups 
of lattices A, B or C), the data on each line are 
constant within $1$ or, in rare cases, 
at most $2$ standard deviations.
In other words,  
there is no statistically significant dependence on the 
lattice spacing in these ratios.

\vskip1ex minus 0.2ex
\noindent
(2)~In all topological sectors 
the results obtained on the lattices B agree
with random matrix theory.
There are deviations
at the level of $2$ or $3$
standard deviations in a few places, 
but this has to be so by the laws of statistics. 

\vskip1ex minus 0.2ex
\noindent
(3)~The results on the lattices C are also
matched by random matrix theory, although in this 
case some differences of up to $4$ standard deviations
are observed at $\nu=1$ and $\nu=2$. 
It is still possible that these are caused
by an unlikely statistical fluctuation, particularly so since
they do not appear to follow any obvious systematic trend.

\vskip1ex minus 0.2ex
\noindent
(4)~In the case of the lattices A, on the other hand, the data  
are in clear disagreement with random matrix theory. Since the
ratios are practically independent of the lattice spacing, we conclude that 
we are seeing the onset of finite-volume
effects that are specific to QCD and are not accounted for by
random matrix theory. Such a regime must in any case eventually be
reached when $L$ is set to values below $1\,\fm$.

\vskip1.0ex minus 0.2ex
For illustration the ratios calculated on lattice ${\rm B}_2$
are plotted 
in fig.~2 together with the values predicted by random matrix theory.
This shows rather clearly that the observed matching is non-trivial  
and at a high level of precision.
Evidently expectation values
of products of eigenvalues could also be considered,
but it is our experience that ratios involving these
are obtained with relatively large statistical errors,
and comparisons with random matrix theory are consequently
less compelling. 

\topinsert
\vbox{
\vskip0.0cm
\epsfxsize=12.0cm\hskip0.0cm\epsfbox{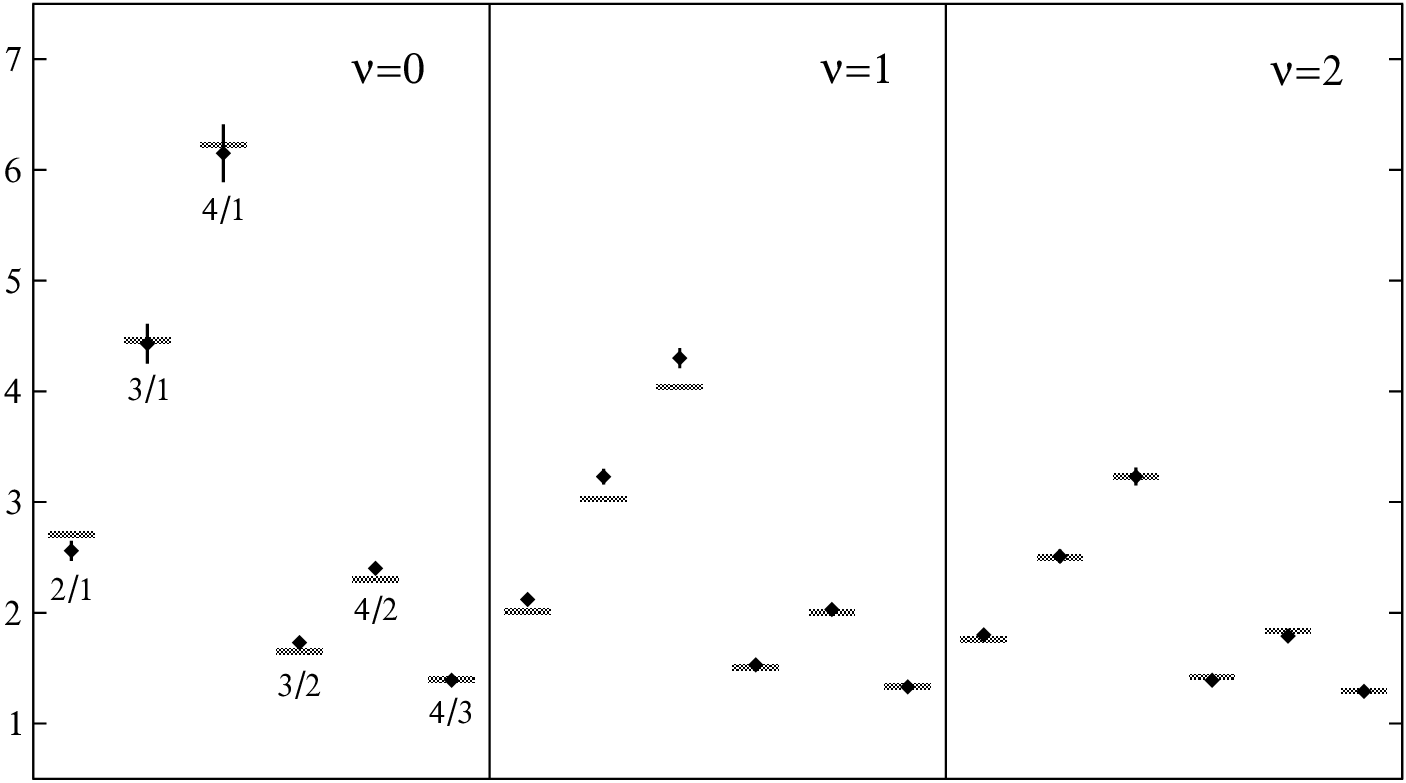}
\vskip0.4cm
\figurecaption{%
Comparison of simulation results for 
$\langle\lambda_k\rangle_{\nu}/\langle\lambda_j\rangle_{\nu}$
from lattice ${\rm B}_2$ (diamonds) with
random matrix theory (horizontal bars) in the sectors with 
topological charge $\nu=0,1,2$.
}
\vskip0.2cm minus 0.2cm
}
\endinsert

\subsection 5.2 Calculation of $\Sigma$

In the present context $\Sigma$ is considered 
to be a parameter with no independent physical meaning
whose value may be fixed by imposing a suitable normalization condition.
We may, for example, require the relation 
\equation{
  \langle \lambda_k\rangle_{\nu}^{\rm QCD}\Sigma V=
  \langle z_k\rangle_{\nu}^{\rm RMT}
  \enum
}
to hold exactly for some $k$ and $\nu$, where $z_k$ denotes
the $k$th scaled eigenvalue in the random matrix model (cf.~subsect.~3.4). 
The numbers obtained in this way
are listed in table~4. 

In the case of the lattices ${\rm A}_1$ and ${\rm A}_2$,
we already know that 
the eigenvalue distributions 
in QCD are poorly matched by those in random matrix theory. The
results for $\Sigma$ consequently show a strong dependence on $k$ and $\nu$,
and any normalization convention that one may adopt is therefore rather 
arbitrary. The situation looks much better on 
the lattices ${\rm B}_0,\ldots,{\rm C}_1$,
where the calculated values of $\Sigma$ are practically
the same for all $k$ and $\nu$ (i.e.~in each column of table~4).
Up to the quoted errors, $\Sigma$ is thus consistently
determined in these cases.

\topinsert
\newdimen\digitwidth
\setbox0=\hbox{\rm 0}
\digitwidth=\wd0
\catcode`@=\active
\def@{\kern\digitwidth}
\tablecaption{Values of $\Sigma r_0^3$ determined from 
$\langle\lambda_k\rangle_{\nu}$}
\vskip0.0ex
$$\vbox{\settabs\+&%
                  xxxx&&
                  xxx&&
                  xxxxxxxxi&&
                  xxxxxxxxi&&
                  xxxxxxxxi&&
                  xxxxxxxxi&i&
                  xxxxxxxxi&i&
                  xxxxxxxxi&i&
                  xxxxxxxxi&i&
                  ii&\cr
\thicktablerule
\vskip1ex
                \+& \hfill $\nu$ \hfill
                 && \hfill $k$\hfill
                 && \hfill ${\rm A}_1$\hfill
                 && \hfill ${\rm A}_2$\hfill
                 && \hfill ${\rm B}_0$\hfill
                 && \hfill ${\rm B}_1$\hfill
                 && \hfill ${\rm B}_2$\hfill
                 && \hfill ${\rm C}_0$\hfill
                 && \hfill ${\rm C}_1$\hfill
                 && \cr
\vskip1.0ex
\thintablerule
\vskip1.5ex
  \+& \hfill $0$\hfill
  &&  \hfill$1$\hfill 
  &&  \hfill $0.215(5)$\hfill
  &&  \hfill $0.242(8)$\hfill
  &&  \hfill $0.221(5)$\hfill
  &&   $@0.274(12)$\hfill
  &&   $@0.280(13)$\hfill
  &&   $@0.263(13)$\hfill 
  &&   $@0.332(18)$\hfill
  &\cr
\vskip0.3ex
  \+& \hfill $0$\hfill
  &&  \hfill$2$\hfill 
  &&  \hfill $0.254(3)$\hfill
  &&  \hfill $0.287(4)$\hfill
  &&  \hfill $0.220(3)$\hfill
  &&  \hfill $0.271(7)$\hfill
  &&  \hfill $0.297(8)$\hfill
  &&  \hfill $0.257(6)$\hfill 
  &&  $@0.298(9)$\hfill
  &\cr
\vskip0.3ex
  \+& \hfill $0$\hfill
  &&  \hfill$3$\hfill 
  &&  \hfill $0.291(2)$\hfill
  &&  \hfill $0.332(3)$\hfill
  &&  \hfill $0.221(2)$\hfill
  &&  \hfill $0.266(5)$\hfill
  &&  \hfill $0.283(5)$\hfill
  &&  \hfill $0.253(5)$\hfill 
  &&  $@0.302(7)$\hfill
  &\cr
\vskip0.3ex
  \+& \hfill $0$\hfill
  &&  \hfill$4$\hfill 
  &&  \hfill $0.329(2)$\hfill
  &&  \hfill $0.377(2)$\hfill
  &&  \hfill $0.220(2)$\hfill
  &&  \hfill $0.260(3)$\hfill
  &&  \hfill $0.284(4)$\hfill
  &&  \hfill $0.246(4)$\hfill 
  &&  \hfill $0.295(6)$\hfill
  &\cr
\vskip1.0ex
\thintablerule
\vskip1.0ex
  \+& \hfill $1$\hfill
  &&  \hfill$1$\hfill 
  &&  \hfill $0.244(3)$\hfill
  &&  \hfill $0.263(5)$\hfill
  &&  \hfill $0.223(3)$\hfill
  &&  \hfill $0.268(7)$\hfill
  &&  \hfill $0.306(7)$\hfill
  &&  \hfill $0.271(9)$\hfill 
  &&  \hfill $0.290(8)$\hfill
  &\cr
\vskip0.3ex
  \+& \hfill $1$\hfill
  &&  \hfill$2$\hfill 
  &&  \hfill $0.276(2)$\hfill
  &&  \hfill $0.307(3)$\hfill
  &&  \hfill $0.220(2)$\hfill
  &&  \hfill $0.266(4)$\hfill
  &&  \hfill $0.292(4)$\hfill
  &&  \hfill $0.270(4)$\hfill 
  &&  \hfill $0.300(6)$\hfill
  &\cr
\vskip0.3ex
  \+& \hfill $1$\hfill
  &&  \hfill$3$\hfill 
  &&  \hfill $0.315(2)$\hfill
  &&  \hfill $0.352(2)$\hfill
  &&  \hfill $0.219(2)$\hfill
  &&  \hfill $0.264(3)$\hfill
  &&  \hfill $0.287(3)$\hfill
  &&  \hfill $0.262(4)$\hfill 
  &&  \hfill $0.296(5)$\hfill
  &\cr
\vskip0.3ex
  \+& \hfill $1$\hfill
  &&  \hfill$4$\hfill 
  &&  \hfill $0.352(2)$\hfill
  &&  \hfill $0.395(2)$\hfill
  &&  \hfill $0.222(2)$\hfill
  &&  \hfill $0.266(3)$\hfill
  &&  \hfill $0.288(3)$\hfill
  &&  \hfill $0.257(3)$\hfill 
  &&  \hfill $0.282(4)$\hfill
  &\cr
\vskip1.0ex
\thintablerule
\vskip1.0ex
  \+& \hfill $2$\hfill
  &&  \hfill$1$\hfill 
  &&  \hfill $0.265(4)$\hfill
  &&  \hfill $0.298(7)$\hfill
  &&  \hfill $0.228(3)$\hfill
  &&  \hfill $0.273(7)$\hfill
  &&  \hfill $0.291(8)$\hfill
  &&  \hfill $0.275(7)$\hfill 
  &&  \hfill $0.312(7)$\hfill
  &\cr
\vskip0.3ex
  \+& \hfill $2$\hfill
  &&  \hfill$2$\hfill 
  &&  \hfill $0.301(3)$\hfill
  &&  \hfill $0.340(4)$\hfill
  &&  \hfill $0.224(2)$\hfill
  &&  \hfill $0.262(5)$\hfill
  &&  \hfill $0.284(5)$\hfill
  &&  \hfill $0.262(6)$\hfill 
  &&  \hfill $0.301(5)$\hfill
  &\cr
\vskip0.3ex
  \+& \hfill $2$\hfill
  &&  \hfill$3$\hfill 
  &&  \hfill $0.335(2)$\hfill
  &&  \hfill $0.382(3)$\hfill
  &&  \hfill $0.223(2)$\hfill
  &&  \hfill $0.262(4)$\hfill
  &&  \hfill $0.289(4)$\hfill
  &&  \hfill $0.258(4)$\hfill 
  &&  \hfill $0.293(5)$\hfill
  &\cr
\vskip0.3ex
  \+& \hfill $2$\hfill
  &&  \hfill$4$\hfill 
  &&  \hfill $0.374(2)$\hfill
  &&  \hfill $0.425(3)$\hfill
  &&  \hfill $0.226(2)$\hfill
  &&  \hfill $0.267(3)$\hfill
  &&  \hfill $0.291(3)$\hfill
  &&  \hfill $0.253(3)$\hfill 
  &&  \hfill $0.287(4)$\hfill
  &\cr
\vskip1.0ex
\thicktablerule
\vskip1.5ex
\+&\kern1pt%
{\footnotefont\noindent%
The errors quoted in this table do not include the error 
on $r_0/a$}&\cr
}$$
\vskip-2ex plus 2ex
\endinsert

There is, however, no reason to expect 
the dimensionless combination $\Sigma r_0^3$ to be independent
of the lattice spacing and the lattice size.
In particular, at fixed lattice spacing,
$\Sigma r_0^3$ appears to be monotonically increasing with
the volume. Close to the continuum limit, $\Sigma$ presumably renormalizes
like the scalar quark density $S$, so that
\equation{
  Z_S\Sigma r_0^3=h(\mu/\Lambda,\Lambda L)+\rmO(a^2),
  \enum
}
where $Z_S(g_0,a\mu)$ 
denotes a renormalization constant, $\mu$ the renormalization 
mass and $\Lambda$ the associated renormalization-group invariant scale.
The analogy with full QCD suggests this,
and, more importantly, the fact that
the higher-order cumulants of $S$ (which renormalize 
multiplicatively) are directly related to the spectral density of 
the Dirac operator [\ref{LeutwylerSmilga}].

The renormalization factor $Z_S$ connecting the lattice theory
to the $\MSbar$ scheme of dimensional regularization has been 
calculated
to one-loop order of perturbation theory
[\ref{AlexandrouEtAl},\ref{CapitaniGiusti}].
This formula is unfortunately not very useful in the present
context, because it largely underestimates
the true value of the renormalization constant
\equation{
   \left.Z_S\right|_{\mu=2\,\GeV}=
   1.43(11)\quad\hbox{at}\quad\beta=6.0,
   \enum
}
which is obtained from non-perturbative scaling studies 
[\ref{HernandezEtAl}]. As a consequence, and since
$Z_S$ is (for $s=0.4$) currently only known at this value of $\beta$,
we are unable to study the scaling behaviour of $Z_S\Sigma r_0^3$.
More extensive simulations would in any case be required 
for a reliable extrapolation to the continuum limit
(if eq.~(5.2) holds).

\topinsert
\vbox{
\vskip0.0cm
\epsfxsize=10.4cm\hskip0.8cm\epsfbox{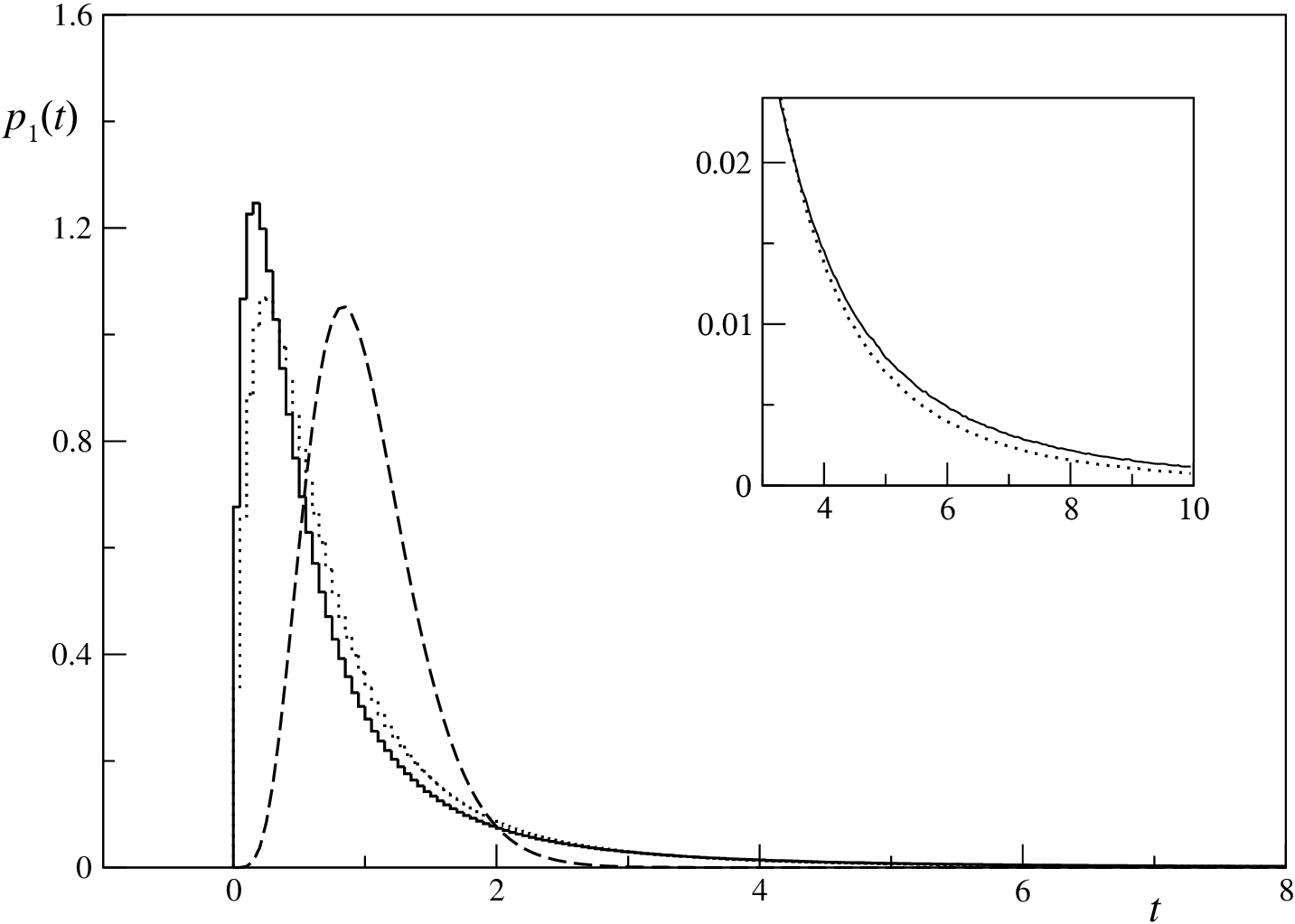}
\vskip0.4cm
\figurecaption{%
Probability distribution $p_1(t)$ of 
the local magnitude $t=\left|\psi_1(x)\right|^2V$ of 
the first norma\-lized eigenvector of $P_{+}D^{\dagger}DP_{+}$
in the charge $Q=0$ sector. 
The solid and dotted lines are the distributions
obtained on the lattices ${\rm C}_1$ and ${\rm B}_2$ respectively,
while the dashed curve is what would be expected for
a random vector with unit norm.
}
\vskip0.0cm
}
\endinsert

\subsection 5.3 Local magnitude of the low modes

In correlation functions of local operators such as the 
pseudo-scalar and the scalar quark
densities, the contribution of the
low modes of the Dirac operator can be large, owing to 
the presence of uncancelled factors of $1/\lambda_k$. 
The actual size of these contributions also depends on the 
magnitude $|\psi_k(x)|^2$ of the associated normalized eigenfunctions 
$\psi_k(x)$ at the points $x$
where the  operators sit.

In fig.~3 we show the distribution of the local magnitude of the 
lowest mode in the vacuum sector. 
Although an average over many gauge configurations is plotted, this
curve is actually fairly universal, i.e.~fluctuations are small
and very similar results are obtained for the higher modes
and in the other topological sectors.
Random matrix theory does not refer to an underlying local structure
and is hence not expected to be relevant here.
In the two chirality sectors, 
the eigenvectors of $\Dhat^{\dagger}\Dhat$ 
are in fact uniformly distributed random vectors,
and the associated distribution (which is also 
displayed in the figure) is clearly different from the one in QCD.

An interesting aspect of the 
QCD distribution shown in fig.~3 is that it decreases only relatively slowly
at large local magnitudes $t$. 
The inset in the figure shows this in greater detail.
In particular, the probability to find a point on the lattice
where $t\geq3$ is more than $5\%$, and there is still a non-negligible
probability of nearly $0.6\%$ for having $t\geq10$.
Correlation functions
of local operators in the $\eps$-regime may consequently
suffer from large statistical fluctuations
that derive from the presence of exceptionally 
low eigenvalues of the Dirac operator in combination with 
accidental ``bumps" in the associated wave functions at the positions of 
the operators.

\section 6. Conclusions

The numerical simulations reported in this paper 
lend further support to the proposition that 
the low-lying eigenvalues of the 
Dirac operator in the $\eps$-regime of QCD are distributed
according to the chiral unitary random matrix model.
Detailed agreement has been observed, in different topological sectors,
on all lattices with linear extent $L$ 
larger than about $1.5\,\fm$. 

Random matrix behaviour sets in rather rapidly
when going from small to large volumes. To some extent
this can be understood by noting that
the average of the lowest eigenvalue in the vacuum sector
is inversely proportional to the volume.
On the lattices ${\rm A}_1$, ${\rm B}_1$ and ${\rm C}_1$,
for example, 
the renormalized spectral gap
\equation{
  \Delta=Z_S^{-1}\langle\lambda_1\rangle_0
  \enum
}
in the $\MSbar$ scheme at $\mu=2\,\GeV$
is approximately equal to $91$, $23$ and $8\,\MeV$ respectively.
Clearly the $\eps$-regime has been safely reached on ${\rm C}_1$,
while this is not so in the case of the lattice ${\rm A}_1$.

Our results also suggest that the lattice theory 
scales coherently to the continuum limit
in all topological sectors (as is usually assumed).
More  extensive simulations
would, however, be needed,
to be able to take the continuum limit 
of the topological susceptibility and other quantities
with confidence.
This is technically possible but requires
significant computer resources, since the 
numerical effort in these calculations 
tends to grow roughly proportionally to $(L/a)^4(L/r_0)^4$.

\vskip1ex
We are indebted to Poul Damgaard, Pilar Hern\'andez, Karl Jansen,
Laurent Lellouch, Ferenc Niedermayer and Giancarlo Rossi
for helpful discussions.
The simulations were
performed on PC clusters at DESY--Hamburg, 
the Institut f\"ur Theo\-retische Physik at the University of Bern,
the Max-Planck-Institut f\"ur Physik in Munich,
the Max-Planck-Institut f\"ur Plasmaphysik in Garching, and
the Leibniz-Rechenzentrum der Bayerischen Akademie der Wissenschaften.
We wish to thank all these institutions 
for supporting our project
and the staff of their computer centres (particularly 
Peter Breitenlohner, Isabel Campos and Andreas Gellrich) for technical help.
L.~G.~was supported in part by the EU under 
contract HPRN-CT-2000-00145 Hadrons/LatticeQCD.

\beginbibliography



\bibitem{GasserLeutwyler}
J. Gasser, H. Leutwyler,
Phys. Lett. B188 (1987) 477;
Nucl. Phys. B307 (1988) 763


\bibitem{LeutwylerSmilga}
H. Leutwyler, A. Smilga,
Phys. Rev. D46 (1992) 5607


\bibitem{ShuryakVerbaarschot}
E. V. Shuryak, J. J. Verbaarschot,
Nucl. Phys. A560 (1993) 306

\bibitem{VerbaarschotZahed}
J. J. Verbaarschot, I. Zahed,
Phys. Rev. Lett. 70 (1993) 3852

\bibitem{Verbaarschot}
J. J. Verbaarschot,
Phys. Rev. Lett. 72 (1994) 2531


\bibitem{VerbaarschotReview}
J. J. Verbaarschot, T. Wettig,
Anu. Rev. Nucl. Part. Sci. 50 (2000) 343

\bibitem{DamgaardReview}
P. H. Damgaard,
Nucl. Phys. B (Proc. Suppl.) 106 (2002) 29


\bibitem{NishigakiDamgaardWettig}
S. M. Nishigaki, P. H. Damgaard, T. Wettig,
Phys. Rev. D58 (1998) 087704


\bibitem{DamgaardNishigaki}
P. H. Damgaard, S. M. Nishigaki,
Phys. Rev. D63 (2001) 045012
[for an important correction see hep-th/0006111]


\bibitem{NumMethods}
L. Giusti, C. Hoelbling, M. L\"uscher, H. Wittig,
Comput. Phys. Commun. 153 (2003) 31

\bibitem{NextPaper}
L. Giusti, P. Hern\'andez, M. Laine, P. Weisz, H. Wittig,
in preparation



\bibitem{GinspargWilson}
P. H. Ginsparg, K. G. Wilson,
Phys. Rev. D25 (1982) 2649


\bibitem{Kaplan}
D. B. Kaplan,
Phys. Lett. B288 (1992) 342;
Nucl. Phys. B (Proc. Suppl.) 30 (1993) 597

\bibitem{Shamir}
Y. Shamir,
Nucl. Phys. B406 (1993) 90

\bibitem{FurmanShamir}
V. Furman, Y. Shamir,
Nucl. Phys. B439 (1995) 54


\bibitem{Hasenfratz}
P. Hasenfratz,
Nucl. Phys. B (Proc. Suppl.) 63 (1998) 53;
Nucl. Phys. B525 (1998) 401

\bibitem{HLN}
P. Hasenfratz, V. Laliena, F. Niedermayer,
Phys. Lett. B427 (1998) 125


\bibitem{NeubergerDirac}
H. Neuberger,
Phys. Lett. B417 (1998) 141;
{\it ibid.}\/ B427 (1998) 353


\bibitem{ExactChSy}
M. L\"uscher,
Phys. Lett. B428 (1998) 342


\bibitem{Locality}
P. Hern\'andez, K. Jansen, M. L\"uscher,
Nucl. Phys. B552 (1999) 363


\bibitem{EdwardsHellerKiskisNarayanan}
R. G. Edwards, U. M. Heller, J. E. Kiskis, R. Narayanan,
Phys. Rev. Lett. 82 (1999) 4188;
Phys. Rev. D61 (2000) 074504

\bibitem{DamgaardEdwardsHellerNarayanan}
P. H. Damgaard, R. G. Edwards, U. M. Heller, R. Narayanan,
Phys. Rev. D61 (2000) 094503

\bibitem{HasenfratzHauswirthEtAl}
P. Hasenfratz, S. Hauswirth, T. J\"org, F. Niedermayer, K. Holland,
Nucl. Phys. B643 (2002) 280

\bibitem{BietenholzEtAl}
W. Bietenholz, K. Jansen, S. Shcheredin,
JHEP 0307 (2003) 033


\bibitem{EvaI}
B. Bunk, K. Jansen, M. L\"uscher, H. Simma,
Conjugate gradient algorithm to compute the low-lying
eigenvalues of the Dirac operator in lattice QCD,
notes (September 1994)

\bibitem{EvaII}
T. Kalkreuter, H. Simma,
Comput. Phys. Commun. 93 (1996) 33


\bibitem{SommerScaleA}
R. Sommer,
Nucl. Phys. B411 (1994) 839

\bibitem{SommerScaleB}
S. Necco, R. Sommer (ALPHA collab.),
Nucl. Phys. B622 (2002) 328


\bibitem{LuigiClaudio}
L. Del Debbio, C. Pica, hep-lat/0309145


\bibitem{AlexandrouEtAl}
C. Alexandrou, E. Follana, H. Panagopoulos, E.~Vicari,
Nucl. Phys. B580 (2000) 394

\bibitem{CapitaniGiusti}
S. Capitani, L.~Giusti,
Phys. Rev. D62 (2000) 114506


\bibitem{HernandezEtAl}
P. Hern\'andez, K. Jansen, L. Lellouch, H. Wittig,
JHEP 0107 (2001) 018;
Nucl. Phys. B (Proc. Suppl.) 106 (2002) 766

\endbibliography

\bye